\documentclass[twocolumn, tighten]{aastex62}
\graphicspath{{./}{figures/}}

\revised{\today}
\submitjournal{ApJL, Accepted on June 14, 2019}

\shorttitle{Submillimeter emission associated with candidate protoplanets}
\shortauthors{Isella et al.}

\begin{document}

\title{Detection of continuum submillimeter emission associated with candidate protoplanets.}

\correspondingauthor{Andrea Isella}
\email{isella@rice.edu}

\author[0000-0002-0786-7307]{Andrea Isella}
\affiliation{Department of Physics and Astronomy, Rice University, 
6100 Main Street, MS-108,
Houston, TX 77005, USA}

\author[0000-0002-7695-7605]{Myriam Benisty}
\affiliation{Departamento de Astronom\'ia, Universidad de Chile, Camino El Observatorio 1515, Las Condes, Santiago, Chile}
\affiliation{Unidad Mixta Internacional Franco-Chilena de Astronom\'ia, CNRS, UMI 3386}
\affiliation{Univ. Grenoble Alpes, CNRS, IPAG, 38000 Grenoble, France.}

\author[0000-0003-1534-5186]{Richard Teague}
\affiliation{Department of Astronomy, University of Michigan, 311 West Hall, 1085 S. University Avenue, Ann Arbor, MI 48109, USA}

\author[0000-0001-7258-770X]{Jaehan Bae}
\affiliation{Department of Terrestrial Magnetism, Carnegie Institution for Science, 5241 Broad Branch Road, NW, Washington, DC 20015,
USA}

\author[0000-0001-7250-074X]{Miriam Keppler}
\affiliation{Max Planck Institute for Astronomy, K\"onigstuhl 17, 69117, Heidelberg, Germany}

\author[0000-0003-4689-2684]{Stefano Facchini}
\affiliation{European Southern Observatory, Karl-Schwarzschild-Str. 2, 85748 Garching, Germany}

\author[0000-0002-1199-9564]{Laura P\'erez}
\affiliation{Departamento de Astronom\'ia, Universidad de Chile, Camino El Observatorio 1515, Las Condes, Santiago, Chile}



\begin{abstract}

We present the discovery of a spatially unresolved source of   
sub-millimeter continuum emission ($\lambda=855$~$\mu$m) associated with 
a young planet, PDS~70~c, recently detected in H$\alpha$ emission around the 5~Myr old 
T Tauri star PDS 70. We interpret the emission as originating from a dusty 
circumplanetary disk with a dust mass between  $2\times10^{-3}$ M$_\Earth$ and $4.2 \times 10^{-3}$ M$_\Earth$. Assuming a standard gas-to-dust ratio of 100, the ratio between 
the total mass of the circumplanetary disk and the mass of the central planet 
would be between $10^{-4}-10^{-5}$. 
Furthermore, we report the discovery of another compact continuum source 
located $0.074\arcsec\pm0.013\arcsec$ South-West of a second known planet 
in this system, PDS~70~b, that was previously detected in near-infrared images. 
We speculate that the latter source might trace dust orbiting in proximity of the planet, but 
more sensitive observations are required to unveil its nature. 
 
\end{abstract}




\section{Introduction} 
\label{sec:intro}

According to current theories, as the mass of a 
forming planet increases above $\sim$10 Earth masses (M$_\Earth$), 
the planet is expected to open a partially 
depleted gap in the circumstellar disk \citep{Crida2006}. Material flowing through the gap 
that enters the region where the gravity of the planet dominates over that of the host 
star, i.e., the planet Hill sphere, is trapped in orbit  
forming a rotating circumplanetary disk (hereafter CPD) \citep{Ward2010}. In analogy with the 
star formation process, circumplanetary material is expected to lose 
angular momentum due to turbulence and viscosity, and accrete onto the planet 
at a rate between $10^{-4}$ and $10^{-8}$ Jupiter masses (M$_J$) per year 
for a period of time comparable with the life time of the circumstellar disk, i.e., 
a few Myr \citep{Lubow2012}. Consequently, for a planet like Jupiter, 
more than 50\% of its mass might have passed through a CPD. 

In addition to regulating the final mass of giant planets, CPDs are the birth place 
of satellites such as the main moons of Jupiter and Saturn. The density,  
temperature, and viscosity of a CPD is expected to control the formation, composition, and architecture 
of the system of satellites \citep{Canup2009}. However, the limited number of moon systems in the Solar system 
and the lack of exo-moon discoveries limit our understanding of how their formation occurs. 
The recent realization that Europa might harbor conditions suitable for life under its icy surface \citep{Pappalardo2013}
stresses the importance to study how moons of giant planets form.

Gas accretion through CPD and the irradiation from both the host star and the central planet 
might warm up gas and dust making the CPD bright at infrared and (sub)millimeter wavelengths \citep{Isella2014, Szu2018, Zhu2018}. 
Furthermore, the gas falling onto the planet and/or its CPD might reach temperatures of thousands of Kelvin 
and emit Hydrogen recombination lines (e.g., H$\alpha$) and UV continuum emission \citep{Zhu2015,marleau17,aoyama18} 

To date, a few giant planet candidates still embedded in their parental circumstellar disks 
have been claimed \citep[e.g.,][]{Kraus2012,Reggiani2018,Sallum2015}, but most still lack confirmation. The most outstanding case is 
the PDS~70 system, which comprises of a $\sim$5 Myr old low mass ($M_\star = 0.76 M_\odot$) T Tauri star at a distance of 
113.4 pc \citep[][and references therein]{Muller2018} surrounded by a disk with a large inner cavity. A planetary-mass companion was detected within the cavity at multiple near-infrared wavelengths ($1.0\mu\textrm{m}<\lambda<3.8\mu \textrm{m}$) and at different epochs between 2012 and 2018 \citep{Keppler2018, Muller2018}.   This companion, PDS 70 b, was also detected in the H$\alpha$ line \citep[$\lambda=0.656$~$\mu$m;][]{Wagner2018}, suggesting that it is accreting gas. Recently, a second companion candidate, PDS 70 c, was discovered in VLT/MUSE observations in the H$\alpha$ line \citep[8-$\sigma$ detection,][]{Haffert2019}. The two planets orbit at about 23~au and 35~au from the central star and 
have masses estimated to range between 4 and 12~M$_J$, while the observed H$\alpha$ line indicates a
mass accretion rate of about $10^{-8}$ M$_J$ yr$^{-1}$. PDS~70 was recently observed at the wavelength 
of 855~$\mu$m with ALMA by \cite{Keppler2019}, who found that most of the circumstellar dust is 
confined in a large dust ring characterized by a radius of 74 au. Furthermore, ALMA observations reveal 
the presence of an inner disk with a radius smaller than 10 au, and a faint spur of dust extending 
from the outer ring toward the inner disk. Interestingly, the latter feature was observed  
at the position of PDS 70~c, though the existence of this planet was not known at the time 
the ALMA data were published. 
Hydrodynamical modelling of the gas kinematics presented in this paper suggested the 
presence of an additional low-mass companion beyond the orbit of PDS 70~b to account 
for the large gap extent.
No emission was reported at the location of PDS 70~b, implying 
an upper limit of about 0.01~$M_\Earth$ for the dust component of a possible CPD.

In Section 2 of this letter, we show that an improved calibration of existing ALMA observations of PDS~70 
reveals that the dust spur identified by \cite{Keppler2019} is a compact source 
of emission spatially separated from the dust ring and located at 
the position of PDS 70~c. Furthermore, we report the detection of a continuum emission close, but not coincident, to the position of PDS~70 b. We label these two sources PDS~70~c$_{smm}$ and PDS~70~b$_{smm}$, respectively. 
In Section 3, we argue that PDS~70~c$_{smm}$ might probe dust emission from a CPD and, 
by comparison with simple CPD models, we estimate a dust mass between $2\times10^{-3}$ and 
$4.2\times10^{-3}$~M$_\Earth$. 
In Section 4 we discuss the main caveat affecting 
this estimate and speculate about the nature of PDS~70~b$_{smm}$. We conclude by 
summarizing our findings in Section 5.

\section{Observations} \label{sec:obs}

\begin{figure*}[!t]
\centering
\includegraphics[width=1.0\textwidth]{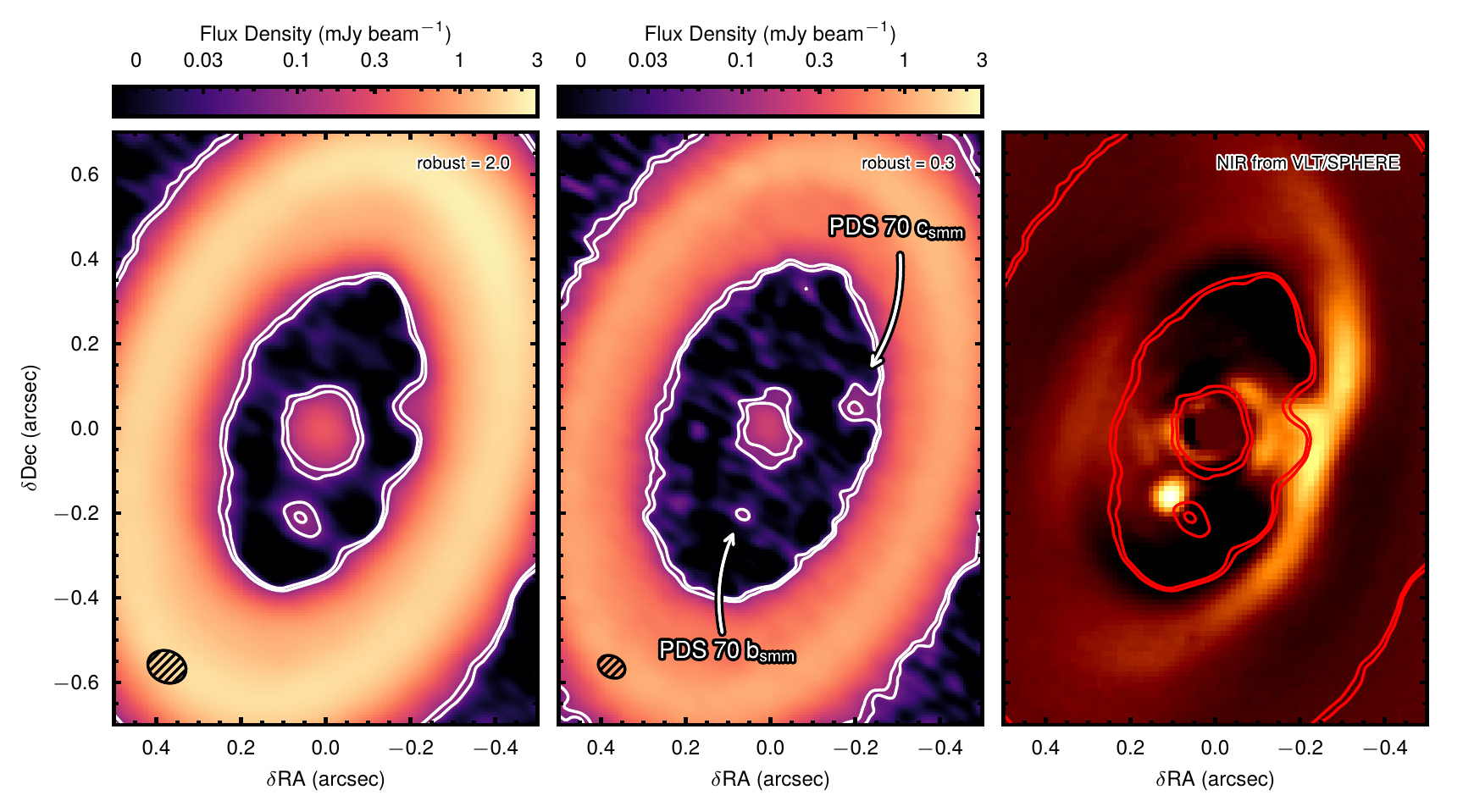}
\caption{ALMA images of the 855 $\mu$m continuum emission recorded 
toward the inner regions of the PDS~70 circumstellar disk. The image on the left was obtained 
adopting a robust parameter of 2 and has an angular resolution (FWHM of the synthesized beam) 
of 0.093\arcsec $\times$ 0.074\arcsec\ indicated by the black ellipse on the bottom left. 
The image in the central panel was created adopting a robust parameter of 0.3 and 
has an angular resolution of 0.067\arcsec $\times$ 0.050\arcsec. White 
contours are drown at 3 and 5 times the rms noise level. See text for the details. 
The labels PDS~70~b$_{smm}$ and PDS~70~c$_{smm}$ indicate the compact sources of 
continuum emission discovered inside the dust cavity. The image on the right shows an 
overlay between the VLT/SPHERE NIR image published in \cite{Muller2018}, and 
the ALMA low intensity red contours from the image with $robust=2.0$  \label{fig:cont}} 
\end{figure*}

The results presented in this letter are based on an improved calibration of the 
ALMA band 7 ($\lambda=855$~$\mu$m) observations of PDS~70 
published in \cite{Keppler2019} and \cite{Long2018}. The source was initially observed in August 2016 
(project ID:2015.1.00888.S) using an array configuration characterized by 
baselines extending between 15~m and 1.5 km, and it was observed again between the 2$^{nd}$ and 
the 6$^{th}$ of December 2017, (project ID:2017.A.00006.S) using a more extended array configuration with 
baselines between 15 m and 6.9 km. The combination of the two sets of observations probes 
angular scales between about 0.025\arcsec\ and 12\arcsec, corresponding to spatial scales 
between 2.8 au and 1360 au at the distance of PDS 70. 
The ALMA correlator was configured to observe both continuum and molecular line emission, but here 
we focus on the continuum emission. We refer to \cite{Keppler2019} for a discussion of the 
observed CO emission. 

Observations taken at the two different epochs were calibrated using the ALMA pipeline and imaged by us 
using the procedure described in \cite{Andrews2018}. One key step of the imaging process is 
the self-calibration of the continuum visibilities which allows to correct short time-scale phase 
fluctuations and improve the imperfect long-time scale phase calibration derived from 
observations of a nearby point-source. Differently from \cite{Keppler2019}, we find 
that self-calibration does improve the continuum image and results in a 40\% increase in 
the peak signal-to-noise ratio (snr). 
Furthermore, in imaging the ALMA data we discovered that the observations acquired in 2016 were 
calibrated by the ALMA pipeline using an incorrect gain calibrator flux, 
which resulted in overestimating the flux density of the PDS~70 disk by about 25\%. 
The self-calibration and imaging was performed in CASA 5.1.1 following the procedure presented in Appendix A. 

We imaged both single epoch and the combined ALMA data using Briggs weighting with {\it robust} 
parameters varying from 2 (which mimics natural weighting) to -0.3 (lower values result in 
significantly higher noise). In Figure~\ref{fig:cont}, we show images of the PDS~70 disk 
obtained from the combined data using $robust=2$ and 0.3. These maps  most clearly reveal 
the presence of faint substructures in continuum emission. However, we stress that our 
results do not depend on the assumed weighting scheme, and, in particular, that the compact sources 
of continuum emission discussed below were identified with the adopted range of weightings in both the combined and 2017 data alone. 

The map generated using $robust=2$ has an rms noise of 18 $\mu$Jy beam$^{-1}$ and achieves 
an angular resolution (FWHM of the synthesized beam) of 0.093\arcsec $\times$ 0.074\arcsec.
In the following, we will refer to this image as the {\it natural} image. Conversely, the map 
generated with $robust=0.3$ (i.e., the {\it robust} image) has a slightly higher noise (rms=19 $\mu$Jy beam$^{-1}$) but 
achieves better angular resolution (FWHM=0.067\arcsec $\times$ 0.050\arcsec). 
Overall, the natural and robust images show the same disk morphology that agrees with the results 
presented in \cite{Keppler2019}. 

The main feature is a bright elliptical ring with semi-major axis of about 0.65\arcsec, semi-minor axis of 0.4\arcsec, position angle of the semi-major axis of about 158\arcdeg\ (as measured from North toward East), and a flux density 
of 177 mJy. Assuming that the dust ring is intrinsically circular, the measured aspect ratio implies a disk inclination of 52\arcdeg. 
The continuum intensity varies along the ring. In the natural image, the intensity reaches a maximum of 2.5 mJy beam$^{-1}$ at a position angle of 325\arcdeg\ and a minimum of 1.7 mJy beam$^{-1}$ at a position angle of 94\arcdeg. A similar relative variation in measured in the robust image. 
A second feature of the continuum emission is a central component characterized by 
a flux density of 0.7 mJy. A 2D elliptical Gaussian fitting of the emission returns a beam deconvolved semi-major axis of 0.10\arcsec$\pm$0.01\arcsec, a semi-minor axis of 0.08\arcsec$\pm$0.01\arcsec, and a position angle of 177\arcdeg$\pm$15\arcdeg. The inclination and position angle of the central component appear to be consistent with those of the dust ring. These two features were discussed in details in \cite{Keppler2019} and will not be further analyzed here.

For the rest of the paper, we focus on the discussion of 
two additional sources of continuum emission detected inside the 
dust cavity. The first, labelled PDS~70~b$_{smm}$, has a peak 
intensity of 100$\pm$18~$\mu$Jy~beam$^{-1}$  (snr $\sim5.5$)
and 73$\pm$19~$\mu$Jy~beam$^{-1}$ (snr $\sim3.8$) in the natural 
and robust images, respectively. A 2D Gaussian fit of 
PDS~70~b$_{smm}$ indicates that the source is much smaller 
than the synthesized beam, suggesting a  beam-deconvolved physical 
diameter $\lesssim 4$ au. 
The second source, labeled PDS~70~c$_{smm}$, is better seen in the robust image 
where it is spatially separated from the dust ring (see also Appendix B). 
PDS~70 c$_{smm}$ has a peak intensity of 106$\pm$19~$\mu$Jy~beam$^{-1}$ (snr $\sim5.6$)
and its morphology is also consistent with a point source. 

The astrometric positions of PDS~70~b$_{smm}$ and PDS~70~c$_{smm}$ are calculated 
with respect to the center of the disk which is defined as the center of the 
innermost component of continuum emission as well as the center of rotation 
as inferred from the $^{12}$CO J=3-2 line emission. These measurements, which are 
discussed in Appendix C, lead to consistent results (Table~\ref{tab:astrometry}).
PDS~70~b$_{smm}$ and PDS~70~c$_{smm}$ are both separated by 0.21\arcsec\ from the center 
of the disk and are located at position angles of about 165\arcdeg\ and 283\arcdeg, 
respectively.

\begin{figure}[!t]
\centering
\includegraphics[width=1.0\linewidth, viewport=0 0 300 300, clip=True]{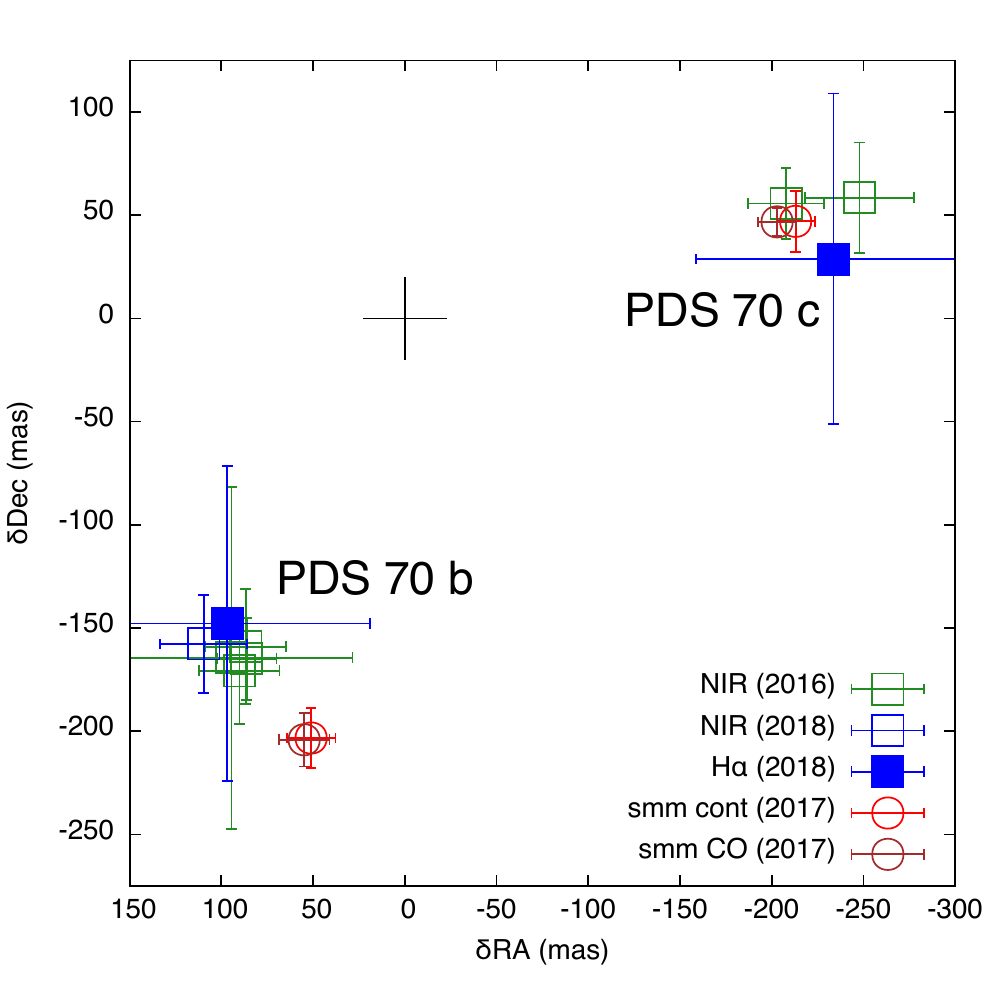}
\caption{Measured astrometry for PDS~70~b and c with 3-$\sigma$ error bars (see Table~\ref{tab:astrometry}). The labels `smm cont' and `smm CO' refer to the position of the ALMA continuum sources relative to the 
center of the continuum emission and the center of the disk rotation as measured from the CO line emission, respectively.
\label{fig:position}}
\end{figure}

\section{Data analysis}

\subsection{Comparing optical, NIR, and sub-mm astrometry}

Figure~\ref{fig:position} compares the relative astrometric positions 
of PDS~70~b$_{smm}$ and PDS~70~c$_{smm}$ to those of  PDS~70~b and PDS~70~c
measured in 2016 and 2018 (see Table 2). 
It must be noted that the position of the ALMA sources are relative to the center 
of the 855~$\mu$m continuum emission and CO kinematics as discussed in Appendix C.
Conversely, the optical and NIR astrometry is measured relative to the stellar position.
In order to compare these measurements, we must therefore assume that the 
star is at the center of the disk.

The position of PDS~70~c$_{smm}$ is in agreement with the optical and NIR 
position of PDS~70~c (see also Figures~\ref{fig:cont} and \ref{fig:complementary}). 
At the estimated orbital radius 34.5 au, the orbital 
period of PDS~70~c should be 230 yr implying an angular yearly motion 
of about 0.008\arcsec\ in the clockwise direction, assuming a circular and coplanar orbit. This is consistent with 
the shift observed between 2016 and 2018.  
PDS~70~b$_{smm}$ appears to be located South-West of PDS~70~b 
at an angular distance of 0.074\arcsec$\pm$0.013\arcsec, as calculated 
by comparing the ALMA position with the closest NIR epoch \citep[2018-02-24 from][]{Muller2018}. 
This offset, which is also 
visible in the overlay between ALMA and VLT/SPHERE observations 
shown in Figure 1, suggests that PDS~70~b$_{smm}$ and PDS~70~b might have different physical origins (see Section 4).  
Although the present analysis has a caveat that the ALMA reference position 
might not correspond to the position of the central star, a substantial offset 
would be required to reconcile the difference and introducing such an offset would lead to 
a disagreement in positions of PDS~70~c and PDS~70~c$_{smm}$. 
Future ALMA observations capable of imaging the innermost disk regions in 
greater details might better pinpoint the position of the central star relative 
to the circumstellar material. 

\subsection{Estimating the CPD dust mass}

The agreement between the position of PDS~70~c$_{smm}$ and its optical/NIR counterpart, 
and the fact that the sub-mm emission is spatially unresolved at the sensitivity and 
resolution of current observations, strongly suggest that PDS~70~c$_{mm}$ traces 
warm dust emission from a CPD. 
This hypothesis is supported by the detection 
of H$\alpha$ line emission attributed to accreting gas \citep[Figure~\ref{fig:complementary},][]{Haffert2019}. 
The planet itself would emit at sub-millimeter wavelengths, 
however we calculate that its 855~$\mu$m flux should be less than 0.1~$\mu$Jy \citep[assuming planet effective temperatures and radii as in][]{Muller2018}.

\begin{deluxetable}{lcccc}
\tabletypesize{\scriptsize}
\tablecaption{Adopted parameters for PDS70~b and PDS70~c \label{table:PDS70}}
\tablehead{
\colhead{Name} & \colhead{M$_p$/M$_J$} & \colhead{$L_p/L_\odot$} & \colhead{a/au} & \colhead{$\dot M/(M_J \, \textrm{yr}^{-1})$}  
}
\colnumbers
\startdata
PDS70 b  & 5--9  & $1.6\times10^{-4}$   &  20.6$\pm$1.2 & $2\times10^{-8}$ \\
PDS70 c  & 4--12 & $1.6\times10^{-4}$   &  34.5$\pm$2.0 & $1\times10^{-8}$ \\
\enddata
\tablecomments{Data from \cite{Wagner2018,Keppler2018,Haffert2019}}
\end{deluxetable}

The spatially integrated flux measured by ALMA can be used to estimate the mass 
of the CPD. Following \cite{Isella2014}, we assume that 
CPDs are vertically isothermal with a radial temperature profile ($T_{CPD}$) controlled by the sum 
of internal viscous heating ($T_{acc})$ and external irradiation from both the central planet ($T_{irr,p}$) 
and the host star ($T_{irr,\star}$),  so that 
\begin{equation}
T_{CPD}^4 = T_{acc}^4 + T_{irr,p}^4 + T_{irr,\star}^4. 
\end{equation}
We assume a stellar luminosity $L_\star=0.36 L_\odot$ \citep{Muller2018}, while 
the physical parameters for PDS~70~c (and b) are listed in Table 1. The luminosity of PDS~70~c 
is not known, but because the masses and ages of PDS~70~b and PDS~70~c are thought to be 
similar, we assume that they might also have similar luminosities. 

The stellar irradiation at the position of PDS~70~c might depend on the geometry of the inner part of the circumstellar disk, as this might 
occult the planets from direct stellar radiation, and on the amount of stellar light 
scattered by the circumstellar dust ring toward the planet itself \citep[see, e.g.][]{Turner2012,Isella2018}. 
In first approximation, a lower limit of $T_{irr,\star}$ might correspond to the midplane 
temperature of an optically thick circumstellar disk, where all the stellar light is absorbed in 
the disk surface and re-processed toward the disk midplane \citep{Chiang1997}. At the orbital 
radius of PDS~70~c, this approximation gives $T_{irr,\star} = 20$~K.
Conversely, an upper limit for  $T_{irr,\star}$ corresponds to the equilibrium temperature 
of a black body embedded in the unattenuated stellar radiation, which gives   
$T_{irr,\star} = 80$~K. 

The temperature due to the irradiation from the planet scales with the distance $r$ as 
\begin{equation}
    T_{irr,p}^4 = \frac{\mu L_p}{4 \pi \sigma r^2 }, 
\end{equation}
where $\mu = 0.1$ is the assumed aspect ratio of the CPD, $L_p$ is the planet luminosity, 
and $\sigma$ is the Stefan-Boltzmann constant. Assuming $L_p=1.6\times 10^{-4}$ L$_\odot$ and $T_{irr,\star}=80$~K, 
we find that $T_{irr,p}>T_{irr,\star}$ at $r<0.1$~au. Instead, if we assume $T_{irr,\star}=20$~K, 
$T_{irr,p}>T_{irr,\star}$ for $r<1.0$~au.

Finally, the temperature due to viscous heating is 
\begin{equation}
T_{acc}^4 = \frac{3GM_p \dot{M}}{8\pi\sigma r^3}\left[ 1- \left( \frac{r_p}{r} \right)^{1/2} \right].  
\end{equation}
For the parameters of PDS~70~c, $T_{acc}$ varies between 1/3 and 1/7 of $T_{irr,p}$, 
suggesting that viscous heating might have a very marginal role. 

\begin{figure}[!t]
\centering
\includegraphics[width=1.\linewidth, viewport=10 10 440 400, clip=True]{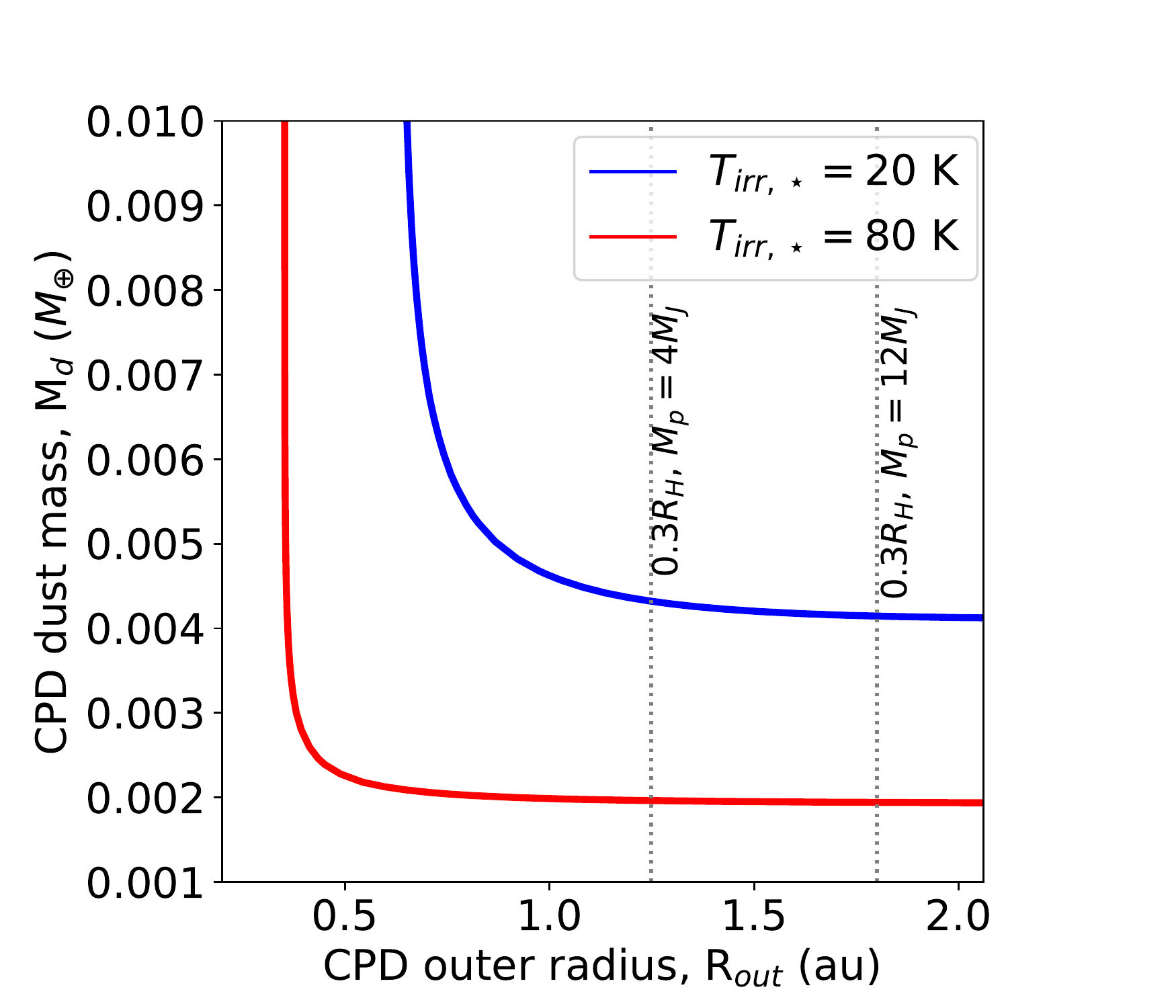}
\caption{Solid lines show the dust mass and outer radius of a CPD characterized by 855~$\mu$m 
continuum fluxes equal to that measured toward PDS~70 c (106~$\mu$Jy). Blue and red colors 
correspond to CPD models where the stellar irradiation alone would result in a CPD temperature 
of 20 K and 80 K, respectively. Vertical dashed lines are drawn at 0.3 times the Hill radius 
for planet masses of 4~M$_J$ and 12~$M_J$. \label{fig:cpdmass}}
\end{figure}

From the CPD temperature derived using Equations 1, 2, 3, and the lower and upper 
limits of $T_{irr,\star}$, we calculate the expected  CPD continuum emission at 
855~$\mu$m using 
Equation 4 in \cite{Isella2014}. To this end, we assume a CPD inclination equal to 
that of the circumstellar disk \citep[$i = 51.7\arcdeg$,][]{Keppler2019}, and a 
dust surface density proportional to $r^{-1}$. \citet{Isella2014} showed that 
the slope of the surface density profile has a little effect on the total disk 
flux. We assume an inner disk radius equal 
to the planet radius, which is estimated to be about 3 Jupiter radii based on the 
planet temperature and luminosity, and a dust 
opacity of 3.5 cm$^{2}$ g$^{-1}$ \citep{Keppler2019}. 
We calculate continuum fluxes for a grid of models characterized by 
dust masses ($M_d$) between $10^{-3}$ M$_\Earth$ and $10^{-1}$ M$_\Earth$, 
and disk outer radii between 0.3~au and 3~au. 
Theoretical models predict that CPDs should be truncated between 1/3 and 1/2 of the planet 
Hill radius ($R_H = a \sqrt[3]{M_p/3M_\star}$) \citep{Quille1998}. For the estimated 
mass range of PDS~70~c, the CPD outer radius might therefore be between 1.4~au and 3~au. 

CPD models that match the 855~$\mu$m continuum flux of PDS~70 c$_{mm}$ 
(106 $\mu$Jy) are shown in Figure~\ref{fig:cpdmass}. As expected, the inferred 
dust mass depends on the disk temperature so that more dust is needed to explain the observed flux if the disk is colder. If $R_{out} \gtrsim 1$ au, the dust mass is 
almost independent on the disk outer radius,  and it reaches values of 
$2\times10^{-3}$~M$_\Earth$ and $4.2 \times 10^{-3}$~M$_\Earth$ in the high and low temperature models, respectively. This behaviour is due to the fact that, in this regime, CPDs are mostly optically thin at 855~$\mu$m. As a consequence, the dust emission 
scales linearly with the dust mass but is independent on the emitting area.
For example, the dust surface density of the CPD model characterized by $T_{irr,\star}=20$~K, $M_d = 4.3\times10^{-3}$~M$_\Earth$, and $R_{out}=1.25$ au ($R_{out} = 0.3 R_H$ for M$_p=4$ M$_J$), is
$\Sigma_d \simeq 0.015 \textrm{g\, cm}^{-2} (\textrm{au}/r)$, which, for the adopted dust 
opacity, corresponds to an optical depth below 1 for $r>0.05$~au. 

If the CPD outer radius is smaller than 1 au, the dust mass required to explain the 
observed emission leads to higher, more optically thick, surface densities.  
In the limit of $\tau \gg 1$, the disk flux scales linearly with the emitting area 
(i.e., $F\propto R_{out}^2$), and iso-flux curves turn from horizontal to vertical. 
The transition between optically thin and thick regime, as well as the minimum outer radius compatible with the observations, depend on the disk temperature. We find that the CPD of 
PDS~70~c might have an outer radius as small as 0.35~au (0.08$R_H$ for M$_p=4$ M$_J$) if $T_{\star,irr}=80$ K, and about 0.6~au (0.15$R_H$) if $T_{\star,irr}=20$ K. 

The same procedure can be applied to estimate the mass and outer radius of a CPD around 
PDS~70~b, although astrometric measurement might suggest that the observed sub-millimeter emission might have a different origin. In this case, the measured 855~$\mu$m flux 
would correspond to dust masses between $1.8 \times 10^{-3}$ M$_\Earth$ and $3.2 \times 10^{-3}$ M$_\Earth$ for optically thin disks larger than $\sim$0.7 au, or a minimum outer radius of 0.2 au if the emission is optically thick.

\section{Discussion}

Inferring the existence and measuring the properties of CPDs is important 
to constrain both planet and moon formation models. The detection of a compact 
source of sub-mm emission at the position of PDS~70~c confirms the hypothesis that 
the H$\alpha$ line emission recently detected by \cite{Haffert2019} probes gas accreting 
onto a young planet from a CPD. The constraints set on the dust mass suggest that the CPD
might have a relatively low mass compared to that of the planet. Using a standard
gas-to-dust ratio of 100, the total mass of the CPD would be $10^{-4}-10^{-5}$ 
times the mass of the planet, unless the CPD is so small ($R_{out} \lesssim 0.1 R_H$) 
to become optically thick. For comparison, the minimum mass of solids required to 
form Jupiter's Galilean moons is $6.5\times10^{-2}$ M$_\Earth$. This corresponds to 
a total gas mass of about 2\% of the mass of Jupiter.  Theoretical models for the formation 
and evolution of giant planets \citep[e.g.][]{Ward2010} predict low CPD/planet mass ratios only 
toward the end of the planet accretion phase, and after the planet has accreted most of 
its mass. 

In discussing the comparison between theory and observations, the caveats related to the
measurement of the CPD mass can not be neglected. Similarly to the case of circumstellar 
disks, the main source of uncertainty in calculating dust masses from sub-millimeter 
fluxes resides in the limited knowledge of the dust opacity. The assumed value of 
3.5 cm$^{2}$ g$^{-1}$ is appropriate for dust grains with chemical composition typical of those 
observed in the interstellar medium and a MRN grain size distribution ($n\propto a^{-3.5}$) 
with a maximum grain size extending a few millimeters \citep{Birnstiel2018}.
However, while millimeter-sized grains have been observed in circumstellar disks, it is 
unclear whether they might exist in a CPD. For example, in the case of PDS~70, it is 
thought that the observed circumstellar dust ring might trap most of the large dust 
grains so that the material flowing from the circumstellar to the circumplanetary disk 
might be depleted in solids, and, in particular, in mm-size grains. Furthermore, mm-size 
grains in the CPD are expected to quickly drift inward as a result of the gas drag
\citep{Zhu2018}. As counter arguments, small solids in the CPD might collide and grow 
in size to form larger particles \citep{Shibaiake2017}, and local maxima in the gas density
could slow down  the inward radial drift of dust particles, as observed in several 
circumstellar disks \citep[see, e.g.,][]{Dullemond2018}. However, if for any reason 
the maximum grain size in the CPD is smaller than about than 100~$\mu$m, the true 
dust opacity might be a factor of several lower than the adopted value, and, consequently, 
the dust mass might be a factor of several larger. Future multi-wavelength ALMA observations 
that measure the spectral slope of the CPD continuum emission will help breaking the 
degeneracy related to the grain size distribution. Furthermore, observations of both optically  thin and thick molecular lines could be used to constrain the gas density and temperature, 
allowing for a more quantitative comparison between observations and models. 

The last item of discussion concerns the nature of PDS~70~b$_{smm}$. Its proximity to 
PDS~70~b suggests that the observed continuum emission might be somehow related to 
the planet. Due to the uncertainties on the position of the host star in the ALMA images, 
we cannot exclude that the sub-millimeter continuum arises for circumplanetary dust. This was 
already discussed in Section 3. 
One possibility to explain the offset might be that PDS~70~b$_{smm}$ traces dust 
particles trapped at the Lagrangian point L5, about $60^\circ$ behind PDS~70~b in azimuth 
along its orbit.
Numerical simulations predict that if the disk viscosity is low ($\alpha < 10^{-4}$) 
solid particles might be trapped at this Lagrangian point for more than 1000 orbits 
\citep{Lyra09, Ricci18, Zhang18}. In the case of PDS~70~b, this would imply particle lifetimes 
larger than about $10^5$ years.
Such a possibility, however, prefers an inclined orbit ($\sim 20^\circ$) with respect to the disk
midplane, assuming that the direction of the ascending node aligns with the circumstellar disk's 
position angle and that optical/NIR and sub-mm emissions arise from the midplane.
Future long-term follow-up observations will better constrain the orbital parameters 
of PDS~70~b and help examine this possibility. 
As an alternative, we speculate it might be possible that optical/NIR traces emission 
from a jet similar to the ones from accreting protostars \citep[e.g.,][]{hartigan11}. 
However, in order to explain the red-shifted H$\alpha$ emission reported by 
\citep{Haffert2019}, the magnetic dipole of the planet must by substantially 
misaligned ($>40\arcdeg$) with respect to the circumstellar disk.
In either case, deep follow-up observations in both optical/NIR and sub-mm wavelengths are highly desired to elucidate the nature of this source.

\section{Conclusions}

We presented the detection of 855~$\mu$m continuum emission at 
the position of the planet PDS~70~c that we attribute to dust emission from a 
CPD. This result supports the hypothesis that this planet is still in its 
accretion phase \citep{Haffert2019}. We have shown that the 
sub-mm flux measured at the position of PDS~70~c constrains the total dust mass 
 to be between  $2\times10^{-3}$ M$_\Earth$ and $4.2 \times 10^{-3}$ M$_\Earth$, if 
the CPD outer radius is larger than about 0.3$R_H$ as predicted by theory. 
Taken at face value, our results indicate that the mass of PDS~70~c CPD  
might be very low compared to that of the planet ($M_d/M_p \sim 10^{-4}-10^{-5}$).
Such low mass ratio and the low mass accretion rate measured by \citet{Haffert2019} suggest 
that the PDS~70~c might have already accreted most of its final mass. 
We reported the detection of another compact source of continuum emission (PDS~70~b$_{smm}$)
located about 0.074\arcsec\ away from PDS~70~b, the other known planet of this system, for which 
we do not currently have any robust interpretation.

The discovery of a CPD at sub-mm wavelengths paves the way for the characterizations 
of the environment surrounding giant planets in the act of forming, and for the study 
of the interaction between the circumstellar and the circumplanetary disk. 
Furthermore, the presented ALMA observations demonstrate the capability to measuring 
orbital parameters of young planets at (sub)-millimeter wavelengths.  
We argue that optical, NIR, and (sub)millimeter observations are highly complementary 
because they probe diverse aspects of planet accretion processes and 
are affected by different systematic errors. The relative astrometric accuracy of ALMA
observations is comparable to that achieved in the optical/NIR and is not contaminated 
by direct or scattered stellar light, which might mislead the interpretation of short 
wavelength observations. However, the offset between PDS~70~b and PDS~70~b$_{smm}$ 
shows that ALMA observations alone might not be sufficient to identify planets in the 
act of forming. As ALMA and existing optical telescopes are reaching their full imaging
capabilities, forthcoming observations of nearby circumstellar disks characterized by 
cavities and gaps like those observed in PDS~70 might reveal more newborn planets 
interacting with their natal disk. Such observations are fundamental to investigating
the processes responsible for the formation of planetary systems.

\acknowledgments
We acknowledge A. Muller, S. Haffert, J. de Boer and A. Vigan for kindly 
sharing their VLT (SPHERE and MUSE) reductions. We tank the anonymous referee 
for helping improving the manuscript.
A. I. acknowledges support from the National Science Foundation
under grant No. AST-1715719. 
M.B. acknowledges funding from ANR of France under contract number ANR-16-CE31-0013 (Planet Forming disks). 
R.T. acknowledges support from NSF grants AST-1514670 and NASA NNX16AB48G.
J.B. acknowledges support from NASA grant NNX17AE31G. 
L.P. acknowledges support from CONICYT project Basal AFB-170002 and from FONDECYT Iniciaci\'on project \#11181068.
M.K. acknowledges funding from the European Union’s Horizon 2020 research and innovation program under grant agreement No 730562 (RadioNet). S.F. acknowledges an ESO Fellowship.

%

\vspace{5mm}
\facilities{ALMA}





\appendix

\section{Self-calibration}

In this section, we provide detailed information about the 
main steps involved in the self-calibration of the ALMA observations of PDS 70. 
The entire self-calibration, starting from the archival data and ending with the 
continuun images presented in this paper, can be reproduced using the python 
script available at \url{http://obelix.rice.edu/~ai14/PDS70/}. The script was 
written for the version 5.1.1 of the Common 
Astronomy Software Applications package (CASA) and was not tested on the more 
recent versions of the software.

\subsection{Self-calibration of 2015.1.00888.S data}

\begin{figure*}[!t]
\centering
\includegraphics[width=0.8\linewidth, viewport=100 0 1500 700 , clip=True]{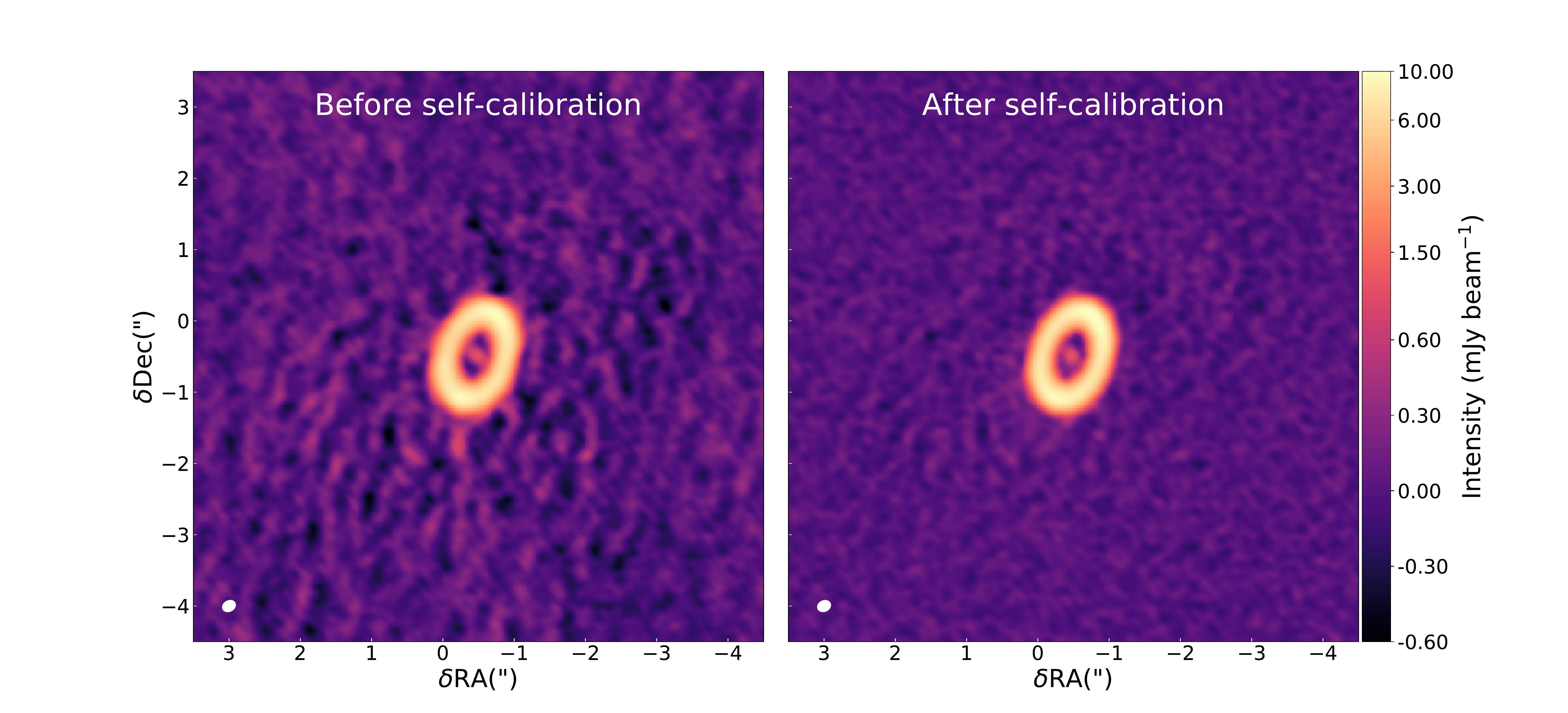}
\caption{Continuum images of the ALMA data obtained as part of project 2015.1.00888.S before (left) and after (right) self-calibration. The color scale has been stretched to highlight the improvement on the image  
noise. 
\label{fig:SB_selfcal}}
\end{figure*}

Following the procedure discussed in \cite{Andrews2018}, we started by calculating antenna based complex phase gains for the data acquired 
in the more compact array configuration, i.e. those from project 2015.1.00888.S. 
Using the task \texttt{tclean}, we generated an image of the continuum
emission adopting Briggs weighting with a robust parameter equal to 0.5, resulting in a 
synthesized beam with a FHWM of 0.19\arcsec$\times$0.15\arcsec (see left panel of Figure~\ref{fig:SB_selfcal}). 
We used multi-scale 
cleaning with scales equal to 0 (point source), 1, 3, and 6 $\times$ the beam FWHM, a cleaning threshold
of 1 mJy beam$^{-1}$ that corresponds to 8$\times$ the rms noise level of the resulting
image, and an elliptical mask with a semi-major axis of 1.7\arcsec, a semi-minor 
axis of 1.1\arcsec, and a position angle of 160\arcdeg.  Setting the cleaning threshold 
to several times the noise level is important to avoid the addition of spurious 
components to the model used to self-calibrate the data, and minimize the effect of 
the adopted mask. We verified that using smaller or larger masks do not affect the results of the self-calibration.  
This initial continuum image has 
a rms noise of 0.13 mJy beam$^{-1}$ and a peak intensity of 10.1 mJy
beam$^{-1}$, corresponding to a peak signal-to-noise ratio (snr) 
of $\sim$78. 
Using the task \texttt{gaincal} we calculated phase gains based on the clean component model generated by \texttt{tclean} on a time 
interval equal to the scan length (\texttt{solint = "inf"}, \texttt{combine=""}). Phase gains were calculated independently for each 
polarization (\texttt{gaintype="G"}) and for each spectral window in 
order to correct for any phase offset between different correlations and spectral bands.
The task \texttt{applycal} was then used to apply the phase gain correction to the data. 
Particular attention should be paid to set the \texttt{spwmap} parameter to properly 
map the spectral phase gain solutions to the corresponding spectral windows of the ALMA data.
This first iteration of phase calibration led to a reduction of about 35\% in the rms noise, and an increase of about 4\% and 70\% in the peak emission and peak snr, respectively.
A second iteration of phase self-calibration was then performed using a 
solution interval of 120 s and averaging on both polarizations
(\texttt{gaintype="T"}) and spectral windows (\texttt{combine="spw"}). 
This led to a further reduction of about 24\% in the rms noise, but a marginal ($\sim$1\%) increase in the peak intensity. The continuum image obtained after the second self-calibration iteration is  shown in the right panel of Figure~\ref{fig:SB_selfcal}. 


 \begin{figure*}[!t]
\centering
\includegraphics[width=1.0\linewidth, viewport=70 100 1550 650, clip=True]{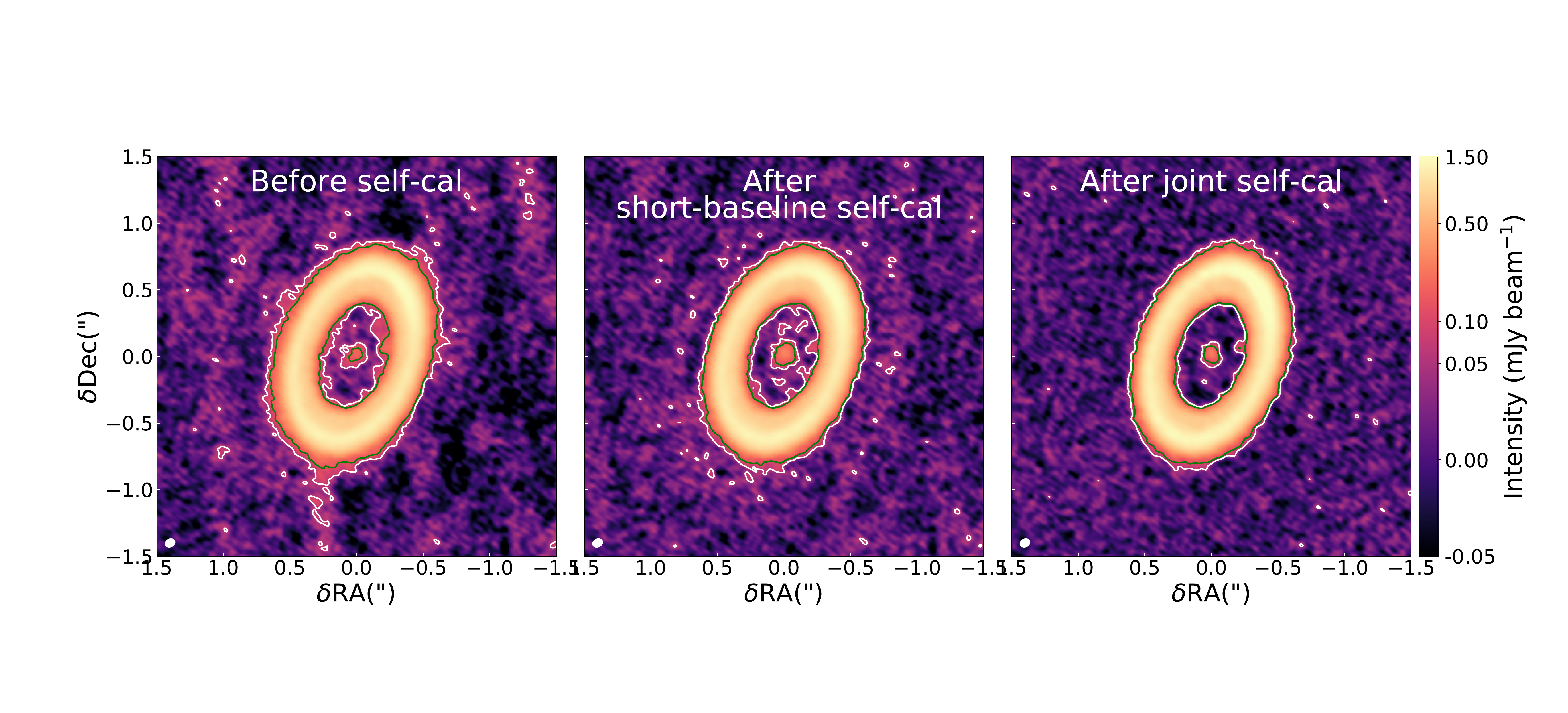}
\caption{Continuum images of the combined ALMA observations of PDS~70 before (left) and after (right) self-calibration. The central panel shows an image obtained after self-calibrating 2015.1.00888.S data and
before self-calibrating 2017.A.00006.S data. The color scale has been stretched to highlight improvement on the image noise. White and green contours are draws at 3$\times$ and 6$\times$ the rms noise levels, which are equal to 0.022 (left), 0.020 (center), 0.018 mJy beam$^{-1}$ (right), respectively. 
\label{fig:comb_selfcal}}
\end{figure*}

 \subsection{Joint self-calibration}
 
 After correcting 2017.A.00006.S data for the source proper motion, we combined them with 
 2015.1.00888.S self-calibrated data and generated a cleaned image using a robust parameter 
 of 0.6, resulting in a beam FWHM of 0.076\arcsec$\times$0.059\arcsec, and a cleaning threshold of 0.15 mJy
 beam$^{-1}$, corresponding to 8$\times$ the rms noise of the resulting image. 
 This initial image of the combined data (central panel of Figure~\ref{fig:comb_selfcal}) has 
 a rms noise level of $\sim$0.020 mJy beam$^{-1}$ and a peak intensity of 1.6 mJy beam$^{-1}$, 
 corresponding to a peak snr of about 77. Using this image, we self-calibrated the combined data by
 averaging both on polarizations and spectral windows adopting solution intervals of 900 s, 120 s 
 and 30 s. The final image (right panel of Figure~\ref{fig:comb_selfcal}) has an rms of 0.018 
 mJy beam$^{-1}$ and a peak intensity of 1.73  mJy beam$^{-1}$, corresponding to a peak snr of 96. 
 Overall, for a robust parameter of 0.6, the self-calibration of the continuum led to a reduction of about 
 of 18\% in the rms noise, and an increase of about 14\% and about 40\% in the peak intensity and peak snr, respectively.  
We found that shorter solution intervals lead to higher noise. We also found that amplitude self-calibration 
 does not provide any improvement of the image quality, and for this reason it was not applied.

\section{Subtraction of a symmetric ring emission}
\begin{figure*}[!t]
\centering
\includegraphics[width=1.0\linewidth, viewport=100 0 1380 600, clip=True]{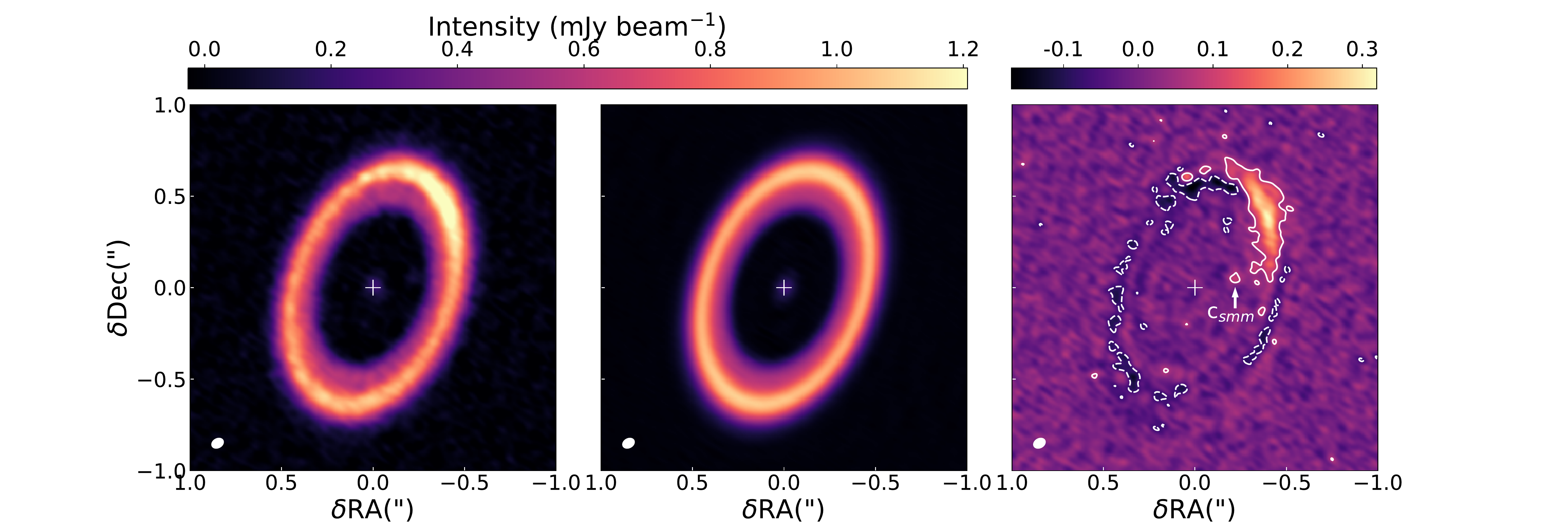}
\caption{Left: Image of the 855~$\mu$m continuum emission recorded toward PDS~70 and 
imaged with a robust parameter equal to 0.3, as in the central panel of 
Figure~\ref{fig:cont}. Center: Image of the symmetric component of the 
continuum emission calculated as discussed in Appendix B. Right: Image of the residuals 
obtained by subtracting the symmetric component of the emission (central panel) from the observations (left panel). 
Dashed and solid contours correspond to $\pm 3 \times$ the 
rms noise level (rms=$19$ $\mu$Jy beam$^{-1}$).
\label{fig:decomposition}}
\end{figure*}

To expose the morphology of PDS~70~c$_{smm}$, 
we subtract the much brighter dust ring and central component from the 
observed continuum emission, assuming that they are symmetric. The subtraction is performed in the 
Fourier space using the following procedure. 
First, we spatially shift the observations to minimize the asymmetry 
of the continuum emission relative to the phase center. This step is 
performed by searching for the phase shift that minimizes the imaginary 
part of the continuum visibilities. The phase shift is defined as 
$\exp[-2\pi i(u \delta\mathrm{RA}+ v \delta\mathrm{Dec})]$, where 
$u$ and $v$ are the spatial frequencies, and $\delta\mathrm{RA}$ 
and $\delta\mathrm{Dec}$ define the translation in the image plane. 
We perform a $\chi^2$ minimization and find a minimum for 
$\delta\mathrm{RA}=0.0119\arcsec$ and $\delta\mathrm{Dec}=0.0116\arcsec$.  
The second step consists in calculating the inclination and position 
angle of the disk that minimize the dispersion of the deprojected visibilities 
calculated on circular annuli in the Fourier space. A $\chi^2$ minimization 
returns a minimum for an inclination of 49.9\arcdeg\, and a position angle of 
159.9\arcdeg. 
The third step consists in deprojecting the observed visibilities using 
the derived inclination and position,  calculating the azimuthally averaged  
profile as a function of the uv-distance, and subtracting it from each 
visibility point.  
Finally, the azimuthally averaged visibilities and the residual visibilities 
are imaged using the same procedure adopted for the observations to map the   
symmetric and asymmetric components of the emission. The images obtained 
using a robust parameter equal to 0.3 are shown in Figure~\ref{fig:decomposition}. 
The image obtained from the residuals of the visibility subtraction more clearly 
shows the crescent along the dust ring and PDS~70~c$_{smm}$, which appears as a 
point source 0.21\arcsec\ away from the center of a disk at a position angle 
of 283\arcdeg. A 2D Gaussian fitting to PDS~70~c$_{smm}$ indicates that the 
emission is not spatially resolved, implying therefore a source 
diameter $\lesssim 4$ au.

\section{Astrometric measurements}
\begin{figure*}[!t]
\centering
\includegraphics[width=0.6\textwidth]{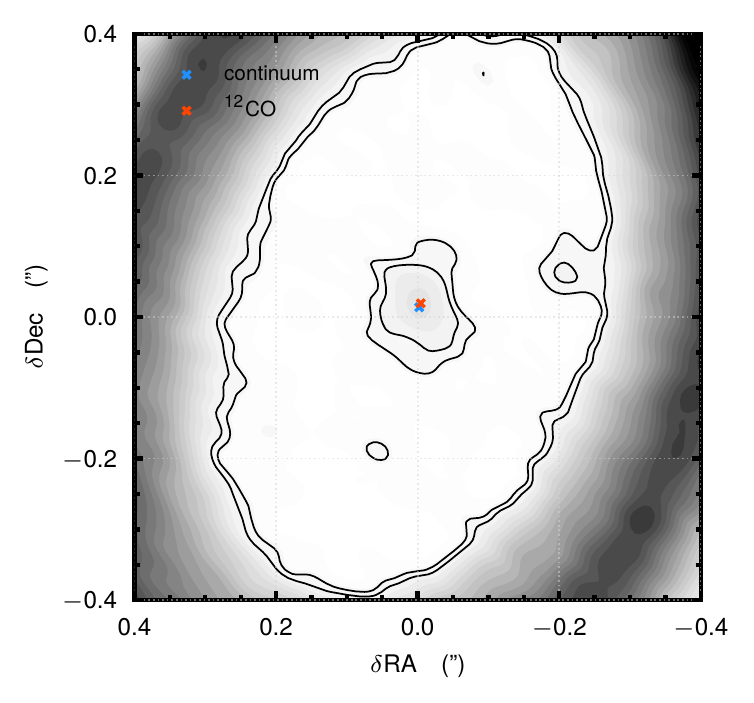}
\caption{Position of the center of the continuum emission (blue cross) and of the disk rotation (red cross). The color scale indicates the continuum intensity. Solid contours are drown at 3 and 5 times the noise level as in the central panel of Figure~\ref{fig:cont}.
\label{fig:centers}} 
\end{figure*}

In Table \ref{tab:astrometry}, we report the relative astrometric measurements of PDS~70~b and c, 
as measured by \citet{Keppler2018}, \citet{Muller2018} and \citet{Haffert2019}, as well as the 
relative position of PDS70~b$_{smm}$ and c$_{smm}$. The latter were calculated using only the 
observations acquired in December 2017 to avoid spatial smearing caused by orbital motion. 
We measure the positions of PDS70~b$_{smm}$ and c$_{smm}$ relative to both the center of the 
disk calculated based on the innermost component of the continuum emission and on the disk rotation 
probed by the $^{12}$CO J=3-2 line emission.
In the first case, we imaged ALMA 2017 continuum data using robust parameters between 2 and 0.6 and measured the 
center of the central component of the continuum emission (see Section 2) through a 
2D Gaussian fit. The average of these measurements defines the center of the continuum emission. 
Since images corresponding to different robust parameters are not independent, the 
uncertainty on the position of the reference point is assumed to be equal to the uncertainty 
on the position of the center as measured in the robust=2 image. The same procedure was 
adopted to calculate the position of PDS~70~b$_{smm}$ and PDS~70~c$_{smm}$. For the latter,  
we only used images generated with robust parameters of 0.6 and 0.7 to avoid 
confusion with the nearby dust ring. 
The center of rotation of the disk is calculated by fitting a Keplerian rotation pattern 
to $^{12}$CO J=3-2 images obtained using natural weighting, as well as robust parameters of 0.7 and 1. 
The CO observations were presented in \cite[][]{Keppler2019} while the fitting procedure is described in 
\cite{Teague2019}. The centers of the continuum emission and of the disk rotation are shown 
in Figure~\ref{fig:centers} and coincide within their uncertainties.

\begin{table*}[!t]
\caption{Relative astrometry of PDS70 b and c. The listed uncertainties correspond 
to a 1$\sigma$ confidence interval. 
 } \label{tab:astrometry}
\begin{center}
\begin{tabular}{ccccccc}
\toprule
Date & Instrument/Filter/$\lambda_0$ ($\mu$m) & $\delta$RA (mas)  & $\delta$Dec  (mas)  & Sep. (mas)  & PA (deg)  & Reference \\
\hline
\multicolumn{7}{c}{{\bf Astrometry of PDS70 b}} \\
2012-03-31 & NICI/L'/3.78 & 58.7$\pm$10.7 & -182.7$\pm$22.2 & 191.9$\pm$21.4 & 162.2$\pm$3.7 & \citet{Keppler2018} \\
2015-05-03 & SPHERE/H3/1.67 & 83.9$\pm$3.6 & -178.5$\pm$4.0  &197.2$\pm$4.0 & 154.9$\pm$1.1 & \citet{Keppler2018} \\
2015-05-31 & SPHERE/H2/1.59 & 89.4$\pm$6.0 & -178.3$\pm$7.1 & 199.5$\pm$6.9  & 153.4$\pm$1.8 & \citet{Keppler2018} \\
2016-05-14 & SPHERE/K1/2.11 & 90.2$\pm$7.3 & -170.8$\pm$8.6 & 193.2$\pm$8.3 & 152.2$\pm$2.3  & \citet{Keppler2018} \\
		   &				 & 	86.2$\pm$5.4  & -164.9$\pm$6.6 & 186.1$\pm$7.0 & 152.4$\pm$1.5 & \citet{Haffert2019} \\
2016-06-01 & NACO/L'/3.80 & 94.5$\pm$22.0 & -164.4$\pm$27.6 & 189.6$\pm$26.3 & 150.6$\pm$7.1 & \citet{Keppler2018} \\
		   &				 & 	86.7$\pm$7.3 	& 	-159.1$\pm$9.3 			& 181.2$\pm$10.0 & 151.4$\pm$2.0 & \citet{Haffert2019} \\
2018-02-24 & SPHERE/K1/2.11 & 109.6$\pm$7.9 & -157.7$\pm$7.9 & 192.1$\pm$7.9 &  147.0$\pm$2.4 & \citet{Muller2018} \\
2018-06-20 & MUSE/H$\alpha$/0.656 & 96.8$\pm$25.9  &  -147.9$\pm$25.4 			& 176.8$\pm$25.0 & 146.8$\pm$8.5 & \citet{Haffert2019} \\
\hline
\multicolumn{7}{c}{{\bf Astrometry of PDS~70~b$_{smm}$}} \\
2017-12-03 & ALMA/855 cont  &  51.1$\pm$4.4 &  -203.3$\pm$4.8 &     -           &      -         &  this work \\
2017-12-03 & ALMA/855 CO    &  54.9$\pm$4.6 &  -204.2$\pm$4.3 &     -           &      -         &  this work \\
\hline
\hline
\multicolumn{7}{c}{{\bf Astrometry of PDS70 c}} \\
2016-05-14 & SPHERE/K1/2.11         & -207.8$\pm$6.9 & 55.7$\pm$5.7 & 215.1$\pm$7.0 & 285.0$\pm$1.5  &\citet{Haffert2019} \\
2016-06-01 & NACO/L'/3.80           & -247.8$\pm$9.9 & 58.5$\pm$8.9 & 254.1$\pm$10.0 & 283.3$\pm$2.0 &\citet{Haffert2019} \\
2018-06-20 & MUSE/H$\alpha$/0.656   & -233.7$\pm$25.0 & 28.8$\pm$26.7 & 235.5$\pm$25.0 & 277.0$\pm$6.5 &\citet{Haffert2019} \\
\hline
\multicolumn{7}{c}{{\bf Astrometry of PDS~70~c$_{smm}$}} \\
2017-12-03 & ALMA/855 cont    & -213.1$\pm$3.5 & 47.0$\pm$4.9 &      -          &        -       &  this work \\  
2017-12-03 & ALMA/855 CO    & -202.9$\pm$4.5 & 46.8$\pm$4.6 &      -          &        -       &  this work \\
\hline
\end{tabular}
\end{center}
\end{table*}

\section{Complementary observations.}
\begin{figure*}[!t]
\centering
\includegraphics[width=0.7\textwidth]{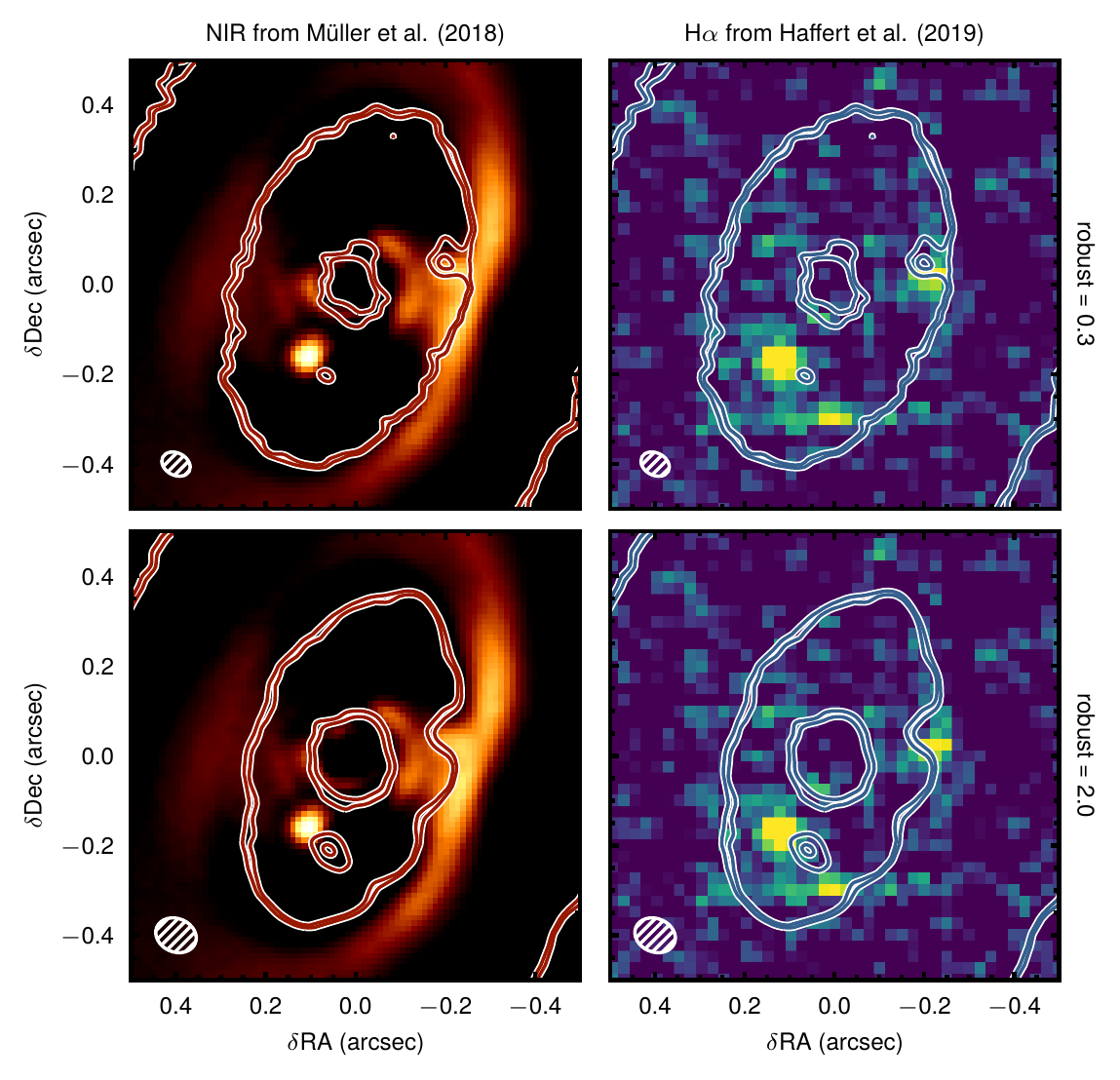}
\caption{VLT/SPHERE NIR image \citep[left,][]{Muller2018} and VLT/MUSE H$\alpha$  image \citep[right,][]{Haffert2019}. The white contours are drawn at 3 and 5 times the rms noise level of the 855 $\mu$m continuum emission, imaged with  $robust=0.3$ (top) and $robust=2$ (bottom). The signal located in the South-West of PDS 70 b in the VLT/MUSE image is likely due to instrumental artifacts \citep{Haffert2019}. 
\label{fig:complementary}} 
\end{figure*}

In Figure\,\ref{fig:complementary}, we show the overlay between the ALMA images of the 855 $\mu$m continuum emission and the VLT/SPHERE NIR image published in \cite{Muller2018} (left panels), and the VLT/MUSE H$\alpha$ image  published in \cite{Haffert2019} (right panels).

\end{document}